\documentclass[structabstract]{sty/aa}

\usepackage[varg]{txfonts}
\usepackage{textgreek}
\usepackage{graphicx}
\usepackage{epstopdf}
\usepackage{subcaption}
\usepackage{txfonts}
\usepackage{pbox}
\usepackage[breaklinks=true,hidelinks]{hyperref}
\usepackage{url}
\usepackage{siunitx}
\usepackage[bottom]{footmisc}
\usepackage[switch]{lineno}
\usepackage{natbib,twoopt}
\usepackage{placeins}  % \FloatBarrier
\makeatletter
\newcommandtwoopt{\citeads}[3][][]{\href{http://adsabs.harvard.edu/abs/#3}%
        {\def\hyper@linkstart##1##2{}%
        \let\hyper@linkend\@empty\citealp[#1][#2]{#3}}}
\newcommandtwoopt{\citepads}[3][][]{\href{http://adsabs.harvard.edu/abs/#3}%
        {\def\hyper@linkstart##1##2{}%
        \let\hyper@linkend\@empty\citep[#1][#2]{#3}}}
\newcommandtwoopt{\citetads}[3][][]{\href{http://adsabs.harvard.edu/abs/#3}%
        {\def\hyper@linkstart##1##2{}%
        \let\hyper@linkend\@empty\citet[#1][#2]{#3}}}
\newcommandtwoopt{\citealtads}[3][][]{\href{http://adsabs.harvard.edu/abs/#3}%
        {\def\hyper@linkstart##1##2{}%
        \let\hyper@linkend\@empty\citealt[#1][#2]{#3}}}
\newcommandtwoopt{\citeyearads}[3][][]%
        {\href{http://adsabs.harvard.edu/abs/#3}
        {\def\hyper@linkstart##1##2{}%
        \let\hyper@linkend\@empty\citeyear[#1][#2]{#3}}}
\newcommandtwoopt{\citeauthorads}[3][][]%
        {\href{http://adsabs.harvard.edu/abs/#3}
        {\def\hyper@linkstart##1##2{}%
        \let\hyper@linkend\@empty\citeauthor[#1][#2]{#3}}}
\makeatother

\begin{document}

   \title{First 230\,GHz VLBI Fringes on 3C~279 using the APEX Telescope}
   \author{
        J. Wagner
                \inst{1,15}
        \and A.L. Roy
                \inst{1}
        \and T.P. Krichbaum
                \inst{1}
        \and W. Alef
                \inst{1}
        \and A. Bansod
                \inst{1}
        \and A. Bertarini
                \inst{1,13}
        \and R. G\"usten
                \inst{1}
        \and D. Graham
                \inst{1}
        \and J. Hodgson
                \inst{1}
        \and R. M\"artens
                \inst{1}
        \and K. Menten
                \inst{1}
        \and D. Muders
                \inst{1}
        \and H. Rottmann
                \inst{1}
        \and G. Tuccari
                \inst{1,5}
        \and A. Weiss
                \inst{1}
        \and G. Wieching
                \inst{1}
        \and M. Wunderlich
                \inst{1}
        \and J.A. Zensus
                \inst{1}
        \and J.P. Araneda
                \inst{2}
        \and O. Arriagada
                \inst{2}
        \and M. Cantzler
                \inst{2}
        \and C. Duran
                \inst{2}
        \and F.M. Montenegro-Montes
                \inst{2}
        \and R. Olivares
                \inst{2}
        \and P. Caro
                \inst{3}
        \and P. Bergman
                \inst{4}
        \and J. Conway
                \inst{4}
        \and R. Haas
                \inst{4}
        \and J. Johansson
                \inst{4,11}
        \and M. Lindqvist
                \inst{4}
        \and H. Olofsson
                \inst{4}
        \and M. Pantaleev
                \inst{4}
        \and S. Buttaccio
                \inst{5}
        \and R. Cappallo
                \inst{6}
        \and G. Crew
                \inst{6}
        \and S. Doeleman
                \inst{6}
        \and V. Fish
                \inst{6}
        \and R.-S. Lu
                \inst{6}
        \and C. Ruszczyk
                \inst{6}
        \and J. SooHoo
                \inst{6}
        \and M. Titus
                \inst{6}
        \and R. Freund
                \inst{7}
        \and D. Marrone
                \inst{7}
        \and P. Strittmatter
                \inst{7}
        \and {L. Ziurys}
                \inst{7}
        \and {R. Blundell}
                \inst{8}
        \and {R. Primiani}
                \inst{8}
        \and J. Weintroub
                \inst{8}
        \and K. Young
                \inst{8}
        \and M. Bremer
                \inst{9}
        \and S. S\'anchez
                \inst{10}
        \and A.P. Marscher
                \inst{12}
        \and R. Chilson
                \inst{14}
        \and K. Asada
                \inst{14}
        \and M. Inoue
                \inst{14}
   }
   \institute{
         % 1
         Max Planck Institute for Radio Astronomy (MPIfR), Bonn, Auf dem H\"ugel 69, DE-53121 Germany \\
             \email{jwagner@kasi.re.kr \; aroy@mpifr.de \; tkrichbaum@mpifr.de}
      \and
         % 2
         European Southern Observatory, Alonso de Cordova 3107, Casilla 19001, Vitacura, Santiago 19, Chile
      \and
         % 3
         Atacama Pathfinder Experiment, Parcela \#85, Sequitor, San Pedro de Atacama, Chile.
      \and
            % 4
%         Onsala Space Observatory (OSO), Chalmers University of Technology, SE-439~92 Onsala, Sweden
Department of Earth and Space Sciences, Chalmers University of Technology, Onsala Space Observatory, 439 92 Onsala, Sweden
       \and
         % 5
         Istituto di Radioastronomia, Istituto Nazionale di Astrofisica, Noto, Italy
       \and
         % 6
         MIT Haystack Observatory, Westford, MA 01886, USA
       \and
         % 7
         Arizona Radio Observatory (ARO), University of Arizona, Tucson, Arizona 85721, USA
       \and
         % 8
         Harvard Smithsonian Center for Astrophysics (CfA), Cambridge, MA 02138, USA
       \and
         % 9
         Institut de Radioastronomie Millim\'{e}trique (IRAM), Saint Martin d'H\`{e}res, France
       \and
         % 10
         Observatorio del Pico Veleta, Estaci\'{o}n RadioAstron\'{o}mica IRAM-IGN, Granada, Spain
       \and
         % 11
         SP Technical Research Institute of Sweden, Bor{\aa}s, Sweden
       \and
         % 12
         Institute for Astrophysical Research, Boston University, Boston, MA 02215, USA
       \and
         % 13
         Institute of Geodesy and Geoinformation, University of Bonn, Germany
       \and
         % 14
         Institute of Astronomy and Astrophysics, Academia Sinica, P.O. Box 23-141, Taipei 10617, Taiwan
       \and
         % 15
         Korea Astronomy and Space Science Institute, 776, Daedeokdae-ro, Yuseong-gu, Daejeon 305-348, Republic of Korea
   }

   \date{2014 February 11 / 2015 June 10} % date of resubmission ; date of submission is 2014 Feb 11 ; date of referee report  2014 Mar 24
      % A&A format: {Received date / Accepted date}
   \abstract
    % Context; leave empty {} if necessary
    {}
    % Aims
    {We report about a 230\,GHz very long baseline interferometry (VLBI) fringe finder observation of blazar 3C~279 with the APEX telescope in Chile, the phased submillimeter array (SMA), and the SMT of the Arizona Radio Observatory (ARO).}
    {We installed VLBI equipment and measured the APEX station position to 1\,cm accuracy ($1\sigma$). We then observed 3C~279 on 2012 May 7 in a 5\,hour 230\,GHz VLBI track with baseline lengths of 2800\,M\textlambda{} to 7200\,M\textlambda{} and a finest fringe spacing of 28.6\,\textmu{as}{}.}
    % Results heading (mandatory)
    {Fringes were detected on all baselines with SNRs of 12 to 55 in 420\,s. The correlated flux density on the longest baseline was $\sim 0.3$\,Jy\ beam$^{-1}$, out of a total flux density of 19.8\,Jy. Visibility data suggest an emission region \mbox{$\lesssim\! 38$\,\textmu{as}{}} in size, and at least two components, possibly polarized. We find a lower limit of the brightness temperature of the inner jet region of about $10^{10}$\,K. Lastly, we find an upper limit of 20\,\% on the linear polarization fraction at a fringe spacing of \mbox{$\sim\!38$\,\textmu{as}{}}.
}
    {With APEX the angular resolution of 230\,GHz VLBI improves to 28.6\,\textmu{as}{}. This allows one to resolve the last-photon ring around the Galactic Center black hole event horizon, expected to be 40\,\textmu{as}{} in diameter, and probe radio jet launching at unprecedented resolution, down to a few gravitational radii in galaxies like M\ 87. To probe the structure in the inner parsecs of 3C~279 in detail, follow-up observations with APEX and five other mm-VLBI stations have been conducted (March 2013) and are being analyzed.}

   %Keywords: http://aas.org/authors/astronomical-subject-keywords-update-august-2013
   \authorrunning{J. Wagner et al.}
   \titlerunning{First 230\,GHz VLBI Observation of 3C~279 using the APEX Telescope}

% Insert title page
\maketitle\

%
%_________________________________________________________________________________________________________________

\section{Introduction} % section 1

Very long baseline interferometry (VLBI) at (sub)millimeter wavelengths offers the unique possibility of high angular resolution studies of objects that are self-absorbed or scatter broadenend at longer
wavelengths. For the nearby supermassive black holes (SMBH) in Sgr\ A$^\star$ and M\ 87 the angular resolution of global mm-VLBI approaches the diameter of the expected black hole shadow and photon ring size \citepads[e.g.,][]{2000ApJ...528L..13F,2015MNRAS.446.1973R}.

In this context a joint effort is being undertaken to build a global (sub)mm-VLBI array to image the immediate environment of nearby BHs.
The angular resolution and imaging capabilities of this array, called the Event Horizon Telescope\footnote{\url{http://www.eventhorizontelescope.org}} (EHT; cf. \citealtads{2008Natur.455...78D,2009astro2010S..68D}), can be improved by adding more stations and longer baselines. As a part of such efforts we carried out a pilot mm-VLBI experiment with the 12\,m APEX telescope\footnote{This publication is based on data acquired with the Atacama Pathfinder Experiment (APEX). APEX is a collaboration between the Max-Planck-Institut f\"ur Radioastronomie, the European Southern Observatory, and the Onsala Space Observatory.} situated near the ALMA array at Chaj\-nan\-tor, Chile. This pilot experiment could be regarded as a pathfinder for the planned participation of ALMA in mm-VLBI.

Two initial attempts at adding a southern mm-VLBI station did not produce fringes (ASTE in 2010 by Honma~et~al. (NAOJ), APEX in 2011). After improvements at APEX in 2012 (for technical details, see \citealt{Roy2013}), we detected the first 230\,GHz fringes at extremely long baselines of 7170\,km to 9450\,km with the currently highest angular resolution of 28.6\,\textmu{as}{}.
This continuation of earlier efforts \citepads[e.g.,][]{2004evn..conf...15K,2008arXiv0812.4211K} enables global 1.3\,mm-VLBI at an angular resolution sufficient to resolve event horizon scale emission in Sgr\ A$^\star$ and M\ 87. These and several other sources will be observed over the next years.

\section{Blazar 3C~279} % section 2
\label{s:source}

The blazar 3C~279 is one of the brightest and best monitored flat-spectrum quasars and was the first object to exhibit apparent superluminal motion. The source of its strong radio to \textgamma{-ray} emission is a relativistic jet of material ejected from near the black hole in 3C~279. The southwest oriented large-scale jet has been extensively studied with $\le\!86$\,GHz VLBI. Adopting the black hole mass of \citetads{2009AA...505..601N}, the redshift of \citetads{1996ApJS..104...37M}, and \textLambda{}CDM cosmology of \citetads{2013ApJS..208...19H}, the scale on 3C~279 is 1.0\,\textmu{as}{}~$\cong$~0.0064\,pc~$\cong$~\num{130}\,$R_\mathrm{s}$.

As one of the brightest quasars 3C~279 is suitable for VLBI fringe finding, although its structure is partially resolved at mm~wavelengths. An observation at 147\,GHz finds a single 34\,\textmu{as}{} component \citepads[but is limited by uv~coverage; cf.][]{2002evn..conf..125K},  whereas 43\,GHz VLBI data from the VLBA-BU-BLAZAR project \citepads{Jorstad2011fermi} suggest three components in the inner 1\,mas. Similarly, in a 230\,GHz EHT observation in 2011 one to two components are seen offset from the core by 80--145\,\textmu{as}{} \citepads{2013ApJ...772...13L}.

%
%_________________________________________________________________________________________________________________

\section{APEX Station Position} % section 2

A high-precision APEX station position is required to minimize the residual fringe rate that causes coherence loss during time integration in the VLBI correlator. Maintaining better than 90\,\% coherence in 1\,s requires the residual fringe rate to be $<\!0.25$\,Hz. At 345\,GHz this corresponds to a position error of $<\!3$\,m.

We derived an initial low-accuracy position (3\,m rms) using a single-band TrueTime XL-AK GPS timing receiver mounted on the telescope, and data logged in 2010 over a one-month period. For better accuracy Onsala later supplied a dual-band Ashtech Micro-Z GPS receiver and TIGO in Concepci\'{o}n lent a choke-ring antenna. We logged dual-band GPS data of the telescope azimuth track over several days in March 2011. Kinematic position solutions were derived to determine the circle center, giving the position of the azimuth axis with 0.3\,mm accuracy. These were adjusted upward by $46 \pm 0.5$\,cm, which is the altitude difference between the reference plane of the GPS antenna and the telescope elevation axis.

The final single- and dual-band GPS station positions agreed to 20\,cm, well within their error bounds. The resulting position of the axis intersection, accurate to 1\,cm ($1\sigma$), is provided below as Geodetic and X, Y, Z coordinates in the ITRF2005 system: \\

\bgroup
\setlength\tabcolsep{4.0pt}
\begin{tabular}{llccl}
Lat. & 23$^{\circ}$00'20.8037"S & ~ & X & \ 2225039.5297\,m \\
Lon. & 67$^{\circ}$45'32.9035"W & ~ & Y & -5441197.6292\,m \\
Alt. & 5104.47\,m               & ~ & Z & -2479303.3597\,m
\end{tabular}
\egroup

%
%_________________________________________________________________________________________________________________

\section{Observation} % section 3

We observed 3C~279 in a three-station 230\,GHz (1.3\,mm) VLBI session on 2012 May 7 %, DOY~128,
from UT\,02:00 to 06:52 during good mm-VLBI weather. The three stations were: 1) the 12\,meter APEX telescope jointly operated by the MPIfR, ESO and OSO on the Chaj\-nan\-tor plateau in Atacama, Chile \citep[see][]{2006A&A...454L..13G}, close to the ALMA array, 2) the Arizona Radio Observatory (ARO) -operated 10\,meter Heinrich Hertz Sub-millimeter Telescope (ARO~SMT) on Mount Graham, Arizona, and 3) the Submillimeter Array (SMA; CfA and ASIAA) on Mauna Kea, Hawaii, with eight 6\,meter dishes phased.  The SMA joined the observation at UT\,03:15.
The observation sampled a 512\,MHz bandwidth (480\,MHz usable) in lower sideband at 229.33\,GHz using DBE1 polyphase filterbank backends (cf. Digital Backend Memo Series, MIT Haystack). At APEX a second backend was installed, the DBBC \citep{doi:10.1117/12.926166}. Data were recorded at 2\,Gbit/s on Mark~5 recorders.
%(cf. VLBA Sensitivity Upgrade Memo\#12, 2008).

\begin{figure}[tb]
        \centering
        \includegraphics[width=0.38\textwidth]{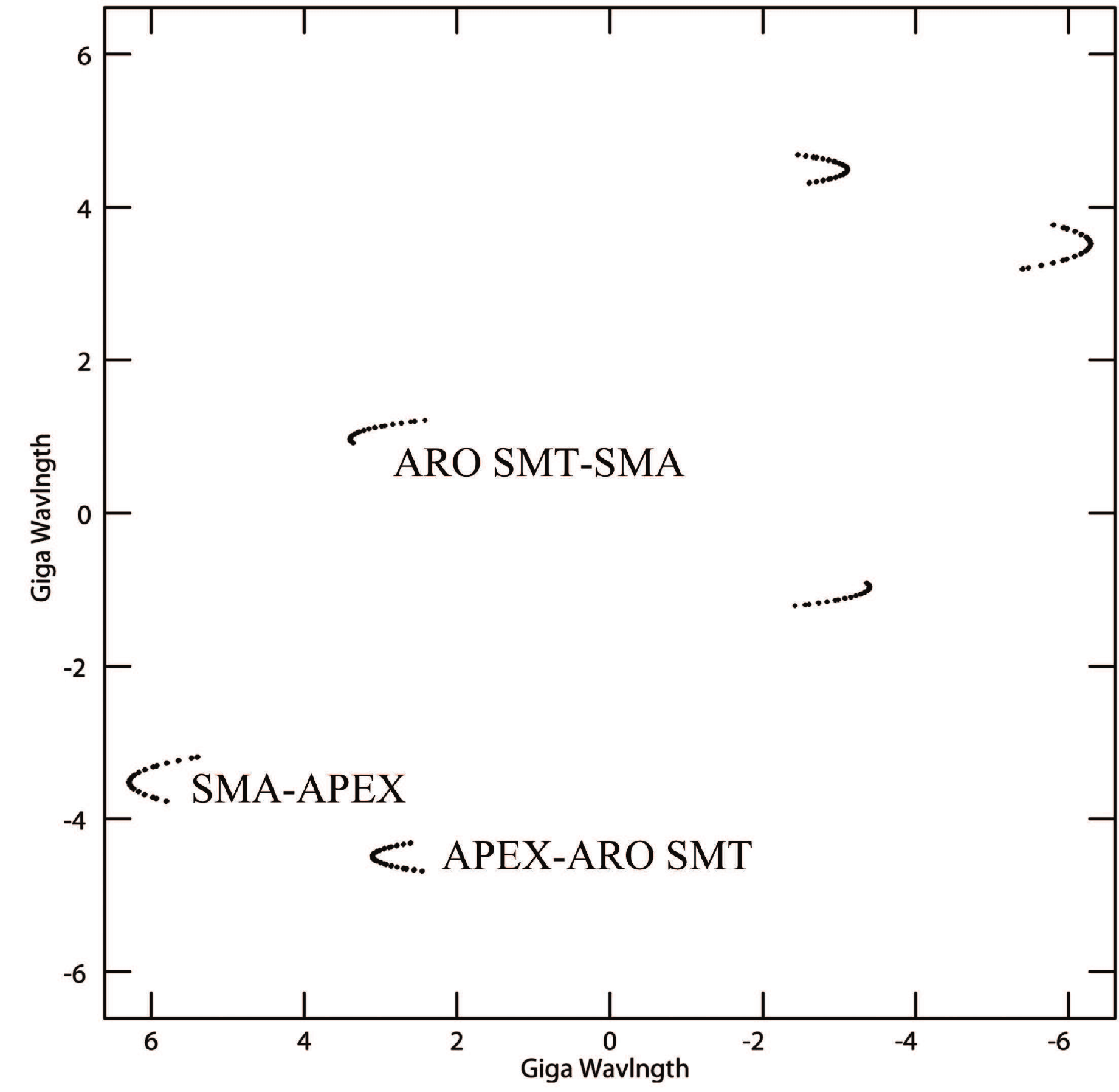}
        \caption{UV coverage on 3C~279 on 2012 May 7 UT 02:00-07:00.}
        \label{fig:uvcoverage}
\end{figure}

The three VLBI baselines (Table~\ref{tab:baselines}) sampled uv~radii of 2800\,M\textlambda{} to 7200\,M\textlambda{}. The shortest fringe spacing was 28.6\,\textmu{as}{}. The uv~coverage on 3C~279 is shown in Fig.~\ref{fig:uvcoverage}.

\begin{table}[tb]
        \caption[APEX 1.3\,mm VLBI fringe spacings and baseline sensitivities.]{Baseline resolution and attained sensitivity} \label{tab:baselines}
        \begin{tabular}{lccccc}
        \hline \hline
        Baseline   & B       & B           & $\theta_\mathrm{B}$ & Obs. time    & $dS_{230}$ \\
        ~          & (km)    & (M\textlambda{}) & (\textmu{as})  & (hours)      & (mJy)      \\ \hline
        SMT--SMA   & $4627$  & $\approx\!3200$      & $58.4$     & 1.75  & 0.07 \\
        APEX--SMT  & $7174$  & $\approx\!5400$      & $37.7$     & 2.33  & 0.18 \\
        SMA--APEX  & $9447$  & $\approx\!6800$      & $28.6$     & 1.75  & 0.08 \\ \hline
        \end{tabular}
        \tablefoot{$\theta_\mathrm{B}$: fringe spacing. $dS_{230}$: median attained sensitivity ($1\sigma$) over a 480\,MHz bandwidth in 10\,s, the shortest atmospheric coherence time.}
\end{table}

%
%_________________________________________________________________________________________________________________

\subsection{Polarization} % section 2.1

The ARO~SMT recorded both circular polarizations. At APEX we used the 230\,GHz single-polarization APEX-1 SHeFI receiver (see \citealt{Belitsky4516518}; \citealtads{2008A&A...490.1157V}) and inserted a \textlambda/4~plate into the optical path to convert linear into circular polarization. The native APEX handedness after 9 mirror reflections (Wieching~2010, private comm.) was LCP (``L''). To maximize experiment output, we switched APEX to RCP (``R'') at UT~04:30 by reorienting the \textlambda/4~plate.
Lastly, the SMA unintentionally observed in a single linear polarization (``X''), thus complicating the interpretation of the resultant visibilities.

%
%_________________________________________________________________________________________________________________

\subsection{Receiver Coherence at APEX} % section 2.3

We monitored the APEX receiver coherence by injecting an H-maser -locked $229$\,GHz pilot tone (PCal) into the SHeFI receiver optics via a horn antenna. The PCal system increased the system temperature negligibly and did not change the nominal APEX sensitivity. % of 0.026\,K/Jy.
The PCal tone phase was measured in 2012, 2013, and again in 2015 with a better tone synthesizer. The receiver coherence ($> 95$\,\% in 1\,s) met the requirements for mm-VLBI.

%_________________________________________________________________________________________________________________

\section{Data reduction} % section 3
\label{s:datareduc}

The station recordings were correlated at the MPIfR Bonn in the DiFX software correlator \citepads{2011PASP..123..275D}. We fringe fitted and reduced the visibility data in NRAO AIPS % 31DEC12
using standard methods. As a cross-check we also reduced the data in the Haystack Observatory Postprocessing System\footnote{\url{http://www.haystack.mit.edu/tech/vlbi/hops.html}} for mm-VLBI (HOPS; cf. \citealtads{1998ASPC..144..407L}), with consistent results.

Calibrations included antenna system temperature data, and a gain % (in kelvin per flux density unit)
calibration on the Moon and Saturn, with a gain estimate used for the phased SMA. Water vapor radiometer and sky dip data were used to correct for atmospheric opacity at a slow cadence. These a~priori calibration data are summarized in Table~\ref{tab:wxtsys}. The amplitude loss due to the SMA linear polarization was compensated by a constant gain of $\sqrt{2}$. We further corrected for DBE1 phase offsets. The optimal fringe fitting interval was 20\,s, twice the shortest coherence time on APEX baselines.
%
%_________________________________________________________________________________________________________________

\begin{table}[tb]
        \caption{Station amplitude calibration data.} \label{tab:wxtsys}
        \centering
        \begin{tabular}{lccccc}
        \hline \hline
        Station & $\tau_{0}$         & $T_\mathrm{sys}$       & Gain           & SEFD           \\
        ~       &  ~                 & (K)                    & (K/Jy)         & (Jy)           \\ \hline
        APEX    & \small{0.06--0.09} & \small{180 (170--185)} & \small{0.0281} &  \small{6410}  \\
        SMT     & \small{0.21--0.33} & \small{250 (230--300)} & \small{0.0182} & \small{13740}  \\
        SMA     & \small{0.03--0.06} & \small{95 (85--125)}   & \small{0.0445} &  \small{2130}  \\ \hline
        \end{tabular}
        \tablefoot{Atmospheric opacities, $\tau_{0}$, were measured with water vapor radiometers. System temperatures, $T_\mathrm{sys}$, % (``mean (min--max)'')
        are double-sideband for ARO~SMT, and an antenna average for the SMA.}
\end{table}

%
%_________________________________________________________________________________________________________________

\begin{figure}[htb]
        \centering
        \includegraphics[width=0.45\textwidth]{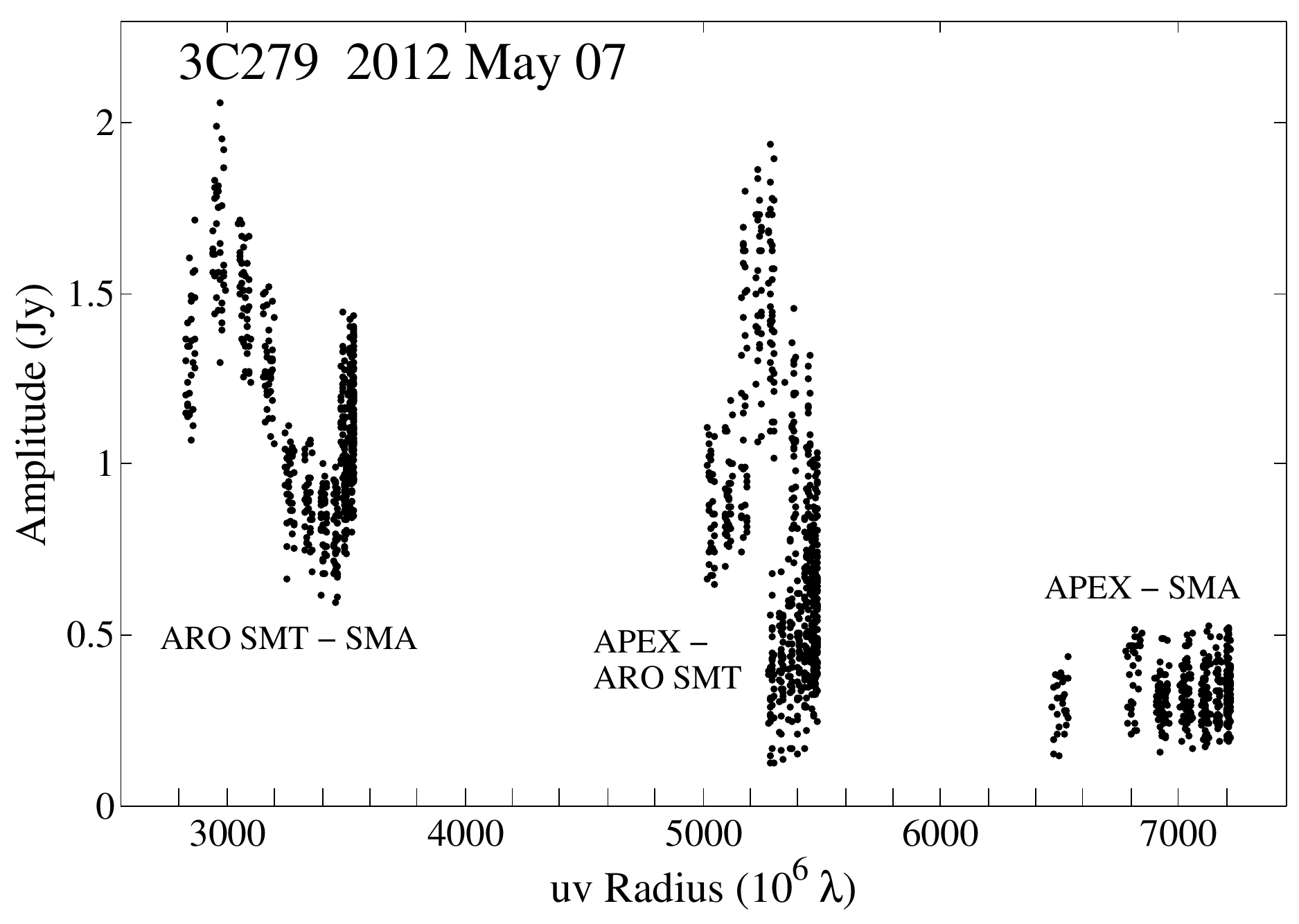}
\vspace{-4px}
        \caption{\label{fig:uvampl} Visibility amplitude (Jy) against uv~distance (M\textlambda{}), averaged using an integration time per point of 10\,s.}
\end{figure}

\begin{figure}[htb]
        \centering
        \includegraphics[width=0.48\textwidth]{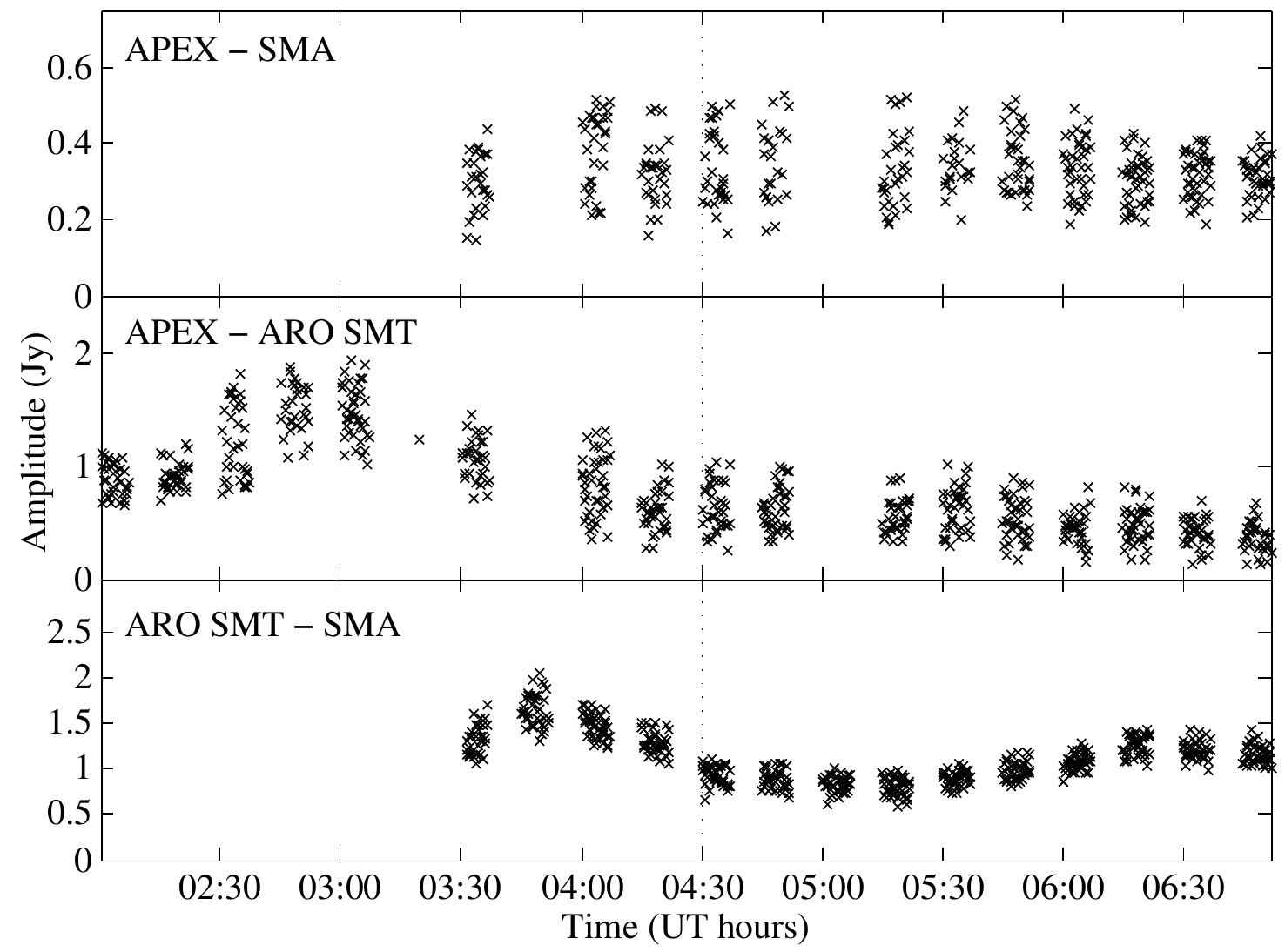}
        \caption{Visibility amplitudes (Jy) against time, averaged to 10\,s per point. The vertical line indicates the polarization swap on the APEX--SMT (LL to RR) and other baselines (LX to RX).\label{fig:baselineampl}}
\end{figure}

\begin{figure}[tbp]
        \centering
        \includegraphics[width=0.48\textwidth]{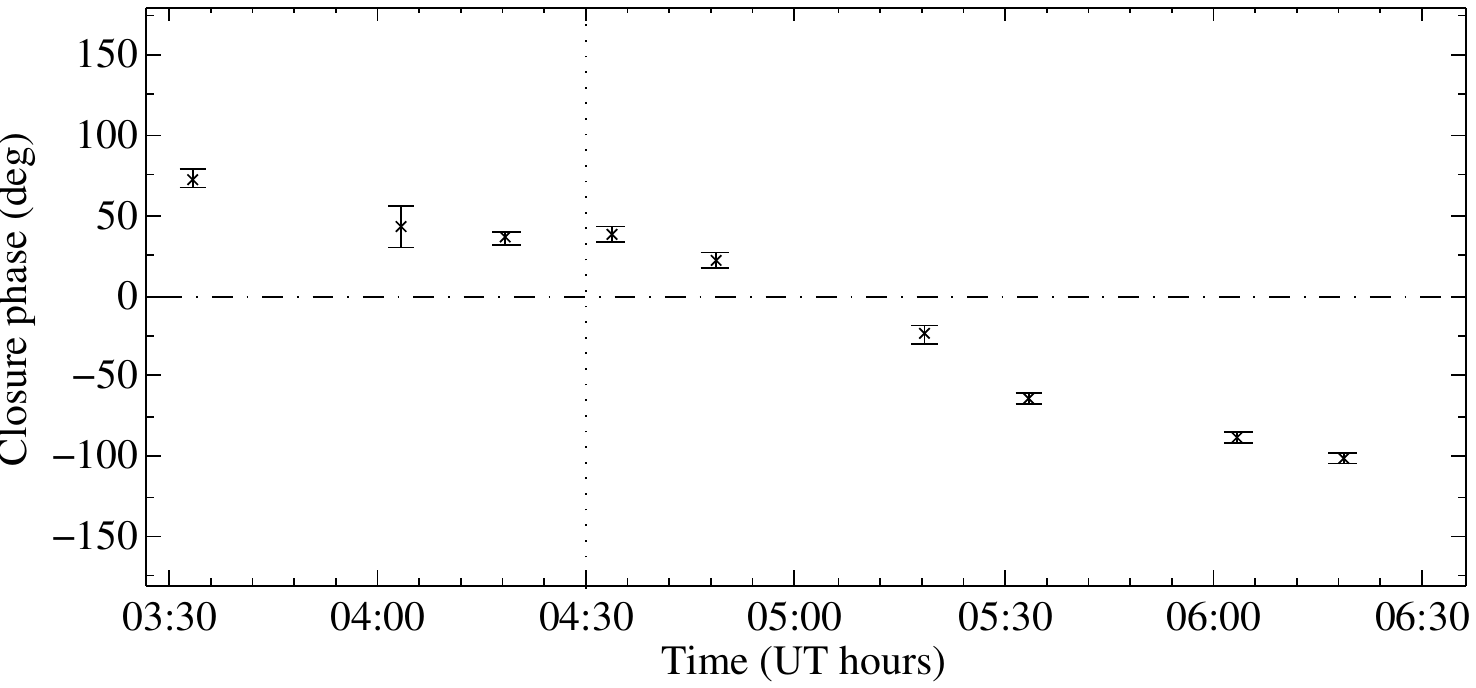}
        \caption{The mixed polarization ``closure phase'' (top) with $1\sigma$ uncertainties, averaged to 420\,s per point. The vertical line indicates the time of polarization change from LLX to RRX. \label{fig:mmreducedifmapcphase}}
\end{figure}

\section{Results and Discussion} \label{s:resultsdiscussion}

Strong fringes were detected on all VLBI baselines. The visibility amplitudes are shown against uv~distance in Fig.~\ref{fig:uvampl}, and against time in Fig.~\ref{fig:baselineampl}. The phased SMA measured a flux density of 19.8\,Jy, agreeing well with the nearest data (2012 April~27, May~11) of their monitoring program \citepads{2007ASPC..375..234G}.
The backends at APEX yielded zero baseline correlation coefficients of \mbox{$>\!0.95$}, demonstrating the DBBC to be ready for mm-VLBI.

The APEX--SMA and ARO\,SMT--SMA baselines produced mixed-hand fringes (LX, RX) with SNRs reported by the HOPS fringe fitter of between 12 and 35 in 420\,s on the former baseline, and SNRs of up to 55 on the latter.
The APEX--ARO\,SMT baseline produced parallel hand fringes (LL, RR) with SNRs between 12 and 35 in 420\,s.
No fringes were detected in the cross-hands (LR, RL). Assuming a detection threshold SNR of 6, this sets an upper limit of 20\,\% on the linear polarization fraction of 3C~279 fine-scale structure. This appears consistent with the nearest VBLA-BU-BLAZAR 43\,GHz maps (2012~April~28, May~26) that show polarization degrees between 1\,\% and 14\,\% for three components detected in the inner 1\,mas of 3C~279 (S. Jorstad, priv. comm.). For other 43\,GHz images, see the VLBA-BU-BLAZAR web page or Jorstad et al. (2015, in preparation).

The mixed-polarization ``closure phase'' on the APEX--ARO\,SMT--SMA triangle (LLX, RRX) is shown in Fig.~\ref{fig:mmreducedifmapcphase}. The closure phase signature of a single source component is zero at all polarization fractions, independent of antenna polarization hands. The detection was not consistent with this signature. This implies two or more components, possibly polarized. Continuity in ``closure phase'' and visibility amplitudes across the APEX polarization swap (Figs.~\ref{fig:baselineampl}, \ref{fig:mmreducedifmapcphase}) further suggests that any polarized sub-structure is only weakly polarized or almost point-like.

Model fitting to constrain the sub-structure is complicated by the mixed polarization setup. It may have corrupted visibilities on SMA baselines and source structure phase in the observed ``closure phase''. These data contain the desired co-polar signal with well studied contaminants (e.g., \citealtads{1996A&AS..116..167M}), as well as cross-polar contamination that is non-trivially dependent  on instrumentation and polarized source structure.
%Briefly, visibilities are the complex sum of LL + RL, where LL is the desired signal and RL is contamination. It is non-trivially dependent on polarized source structure and other effects.
We simulated models of plausible 3C~279 fine-scale structure with different degrees of component polarization using AIPS task DTSIM, modified to generate mixed polarization visibilities.
The closure phase corruption was found to be mostly small relative to the trend in Fig.~\ref{fig:mmreducedifmapcphase}. In contrast, the amplitudes on SMA baselines were impacted by larger polarization-dependent offsets and some parallactic angle -dependent variation. Robustly untangling source structure and contaminations in the observed mixed-polarization data proved challenging.

\begin{figure}[tb]
        \centering
        \includegraphics[width=0.48\textwidth]{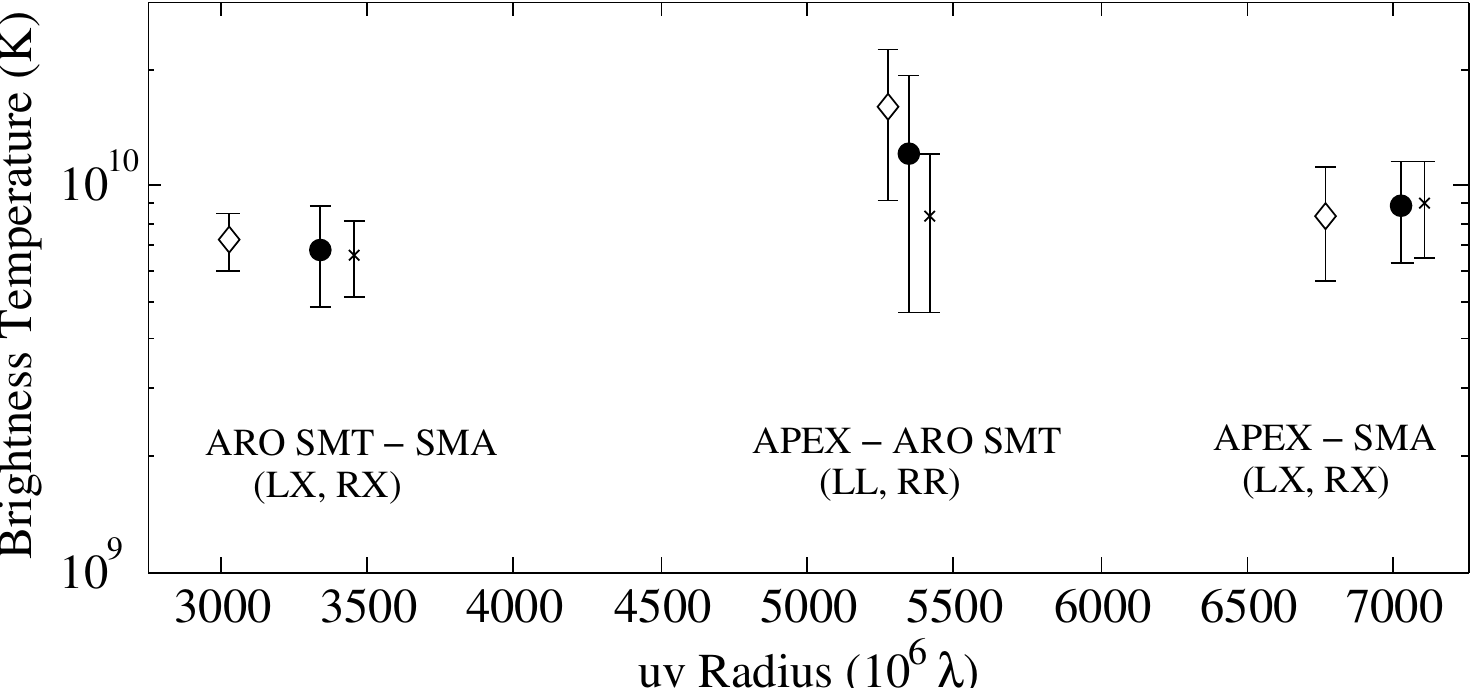}
        \caption{Lower limits of the inner jet brightness temperature based on visibilities before (diamonds), after (crosses), or across (solid) the APEX polarization switch, with $1\sigma$ uncertainties. \label{fig:TbMultifreq}}
\vspace{-0.5cm}
\end{figure}

The APEX--ARO~SMT visibilities (LL, RR) are unaffected by the SMA polarization.
We thus follow a common approach used for sparse visibility data to derive a lower limit of the brightness temperature, $T_\mathrm{b}$, of the inner jet region, based on visibility amplitudes, $S(q)$, against uv~distance, $q$ (see, e.g., \citealtads{2015A&A...574A..84L}). Two assumptions are necessary. Firstly, that visibility amplitudes are dominated by the radio core, and secondly, that the resolved structure is sufficiently well described by a circular Gaussian component ($T_\mathrm{b} = [2 \ln{2} / (\pi \, k_\mathrm{B})] \, S \, \lambda^2/\theta^2$), with a full width at half maximum size, $\theta$. A fringe detection implies that $\theta$ is smaller than the respective fringe spacing, i.e., $\theta \le 1/q$.

The resultant {\em lower limit} of $T_\mathrm{b}$ of the inner jet region is about \mbox{$10^{10}$\,K}. It is shown against uv~distance in Fig.~\ref{fig:TbMultifreq}, with ``lower limits'' on SMA baselines shown for reference.
Lastly, we note that {\em estimating} the $T_\mathrm{b}$ of the core, and identifying other inner jet -related fine-scale structure, requires future quasi-simultaneous VLBI imaging observations at 1.3\,mm and longer wavelengths.

%
%_________________________________________________________________________________________________________________

\section{Conclusions} % section 6

The first APEX 1.3\,mm VLBI fringes with a longest baseline of 7200\,M\textlambda{} and a shortest fringe spacing of 28.6\,\textmu{as}{} demonstrated that new mm-VLBI observations can now be carried out with an improved angular resolution. For Sgr~A$^\star$ and M\ 87 in particular a sufficiently high resolution can now be reached to directly probe the accretion physics and strong relativistic effects at event horizon scales. The current  3C~279 mixed-polarization data at this unprecedented high resolution showed hints of fine-scale structure, but did not allow a detailed study. New 1.3\,mm VLBI observations that include 3C 279 have been made (March 2013) with the largest array yet available at this frequency, now six stations including APEX. These new data are being analyzed.

%
%_________________________________________________________________________________________________________________

\section{Acknowledgments}

We thank Hayo Hase (BKG/TIGO) for his prompt support with the loan of a replacement GPS antenna that enabled us to measure the station position.
We thank Lars-{\AA}ke Nyman for arranging the FTP transfer of test data from the ALMA OSF, and for finding the historical GPS installation of \citetads{2001PCEA...26..421G} at the ALMA site.
We thank Svetlana Jorstad for her input on polarization.
We also thank all people at the telescopes, institutes, and companies for their indispensable help and support that made APEX mm-VLBI possible.
This study makes use of 43 GHz VLBA data from the Boston University gamma-ray blazar monitoring program VLBA-BU-BLAZAR (\url{http://www.bu.edu/blazars/VLBAproject.html}), funded by NASA through the Fermi Guest Investigator Program. VLBI research at the Sub-millimeter Telescope of the Arizona Radio Observatory is partially supported by the NSF University Radio Observatories Program (URO grant AST-1140030) and by AST-0905844. The ARO is a facility of Steward Observatory, University of Arizona.

{\it Facilities:} APEX, SMA, ARO~SMT.  % http://aas.org/aastex/facility-keywords

%%%%%%%%%%%%%%%%%%%%%%%%%%%%%%%%%%%%%%%%%%%%%%%%%%%%%%%%%%%%%%%%%%%%%%%%%%%%%%%%%%%%%%%%%%%%%%%%%%%%%%%%%%%

\bibliography{bib/apex-2012-05-aa-rnote}

\end{document}